\documentclass[10pt,journal,compsoc]{IEEEtran}
\ifCLASSOPTIONcompsoc
  \usepackage[nocompress]{cite}
\else
  \usepackage{cite}
\fi
\usepackage{amsthm}
\usepackage{graphicx} 
\usepackage{amsmath}
\usepackage{amssymb}
\usepackage{subfigure}
\usepackage{epstopdf}
\usepackage{algorithm}
\usepackage{algpseudocode}
\usepackage{multirow}
\usepackage{array}

\usepackage{color}
\usepackage{colortbl}
\usepackage{ragged2e}
\usepackage{hyperref}
\usepackage{cite}
\usepackage{booktabs}
\usepackage{amsmath}
\usepackage{comment}
\DeclareMathOperator*{\argmin}{argmin} 
\ifCLASSINFOpdf
\else
\fi
\begin{document}
%
\title{Constant Sequence Extension for \\ Fast Search Using Weighted Hamming Distance}
%
%
%
%

\author{Zhenyu~Weng,~\IEEEmembership{Member,~IEEE,}
	Huiping~Zhuang,~\IEEEmembership{Member,~IEEE,}
	Haizhou~Li,~\IEEEmembership{Fellow,~IEEE,}
	Zhiping~Lin,~\IEEEmembership{Senior~Member,~IEEE,}


	\thanks{Zhenyu Weng and Zhiping Lin are with School of Electrical and Electronic Engineering,
	Nanyang Technological University, Singapore 639798 (e-mail: zhenyu.weng@ntu.edu.sg and ezplin@ntu.edu.sg)(Corresponding Author: Zhiping Lin).}
\thanks{Huiping Zhuang is with Shien-Ming Wu School of Intelligent Engineering, South China University of Technology, Guangdong, China 511442 (email: hpzhuang@scut.edu.cn)}
\thanks{Haizhou Li is with Shenzhen Research Institute of Big Data, School of Data Science, The Chinese University of Hong Kong, Shenzhen, China 518172 (email: haizhouli@cuhk.edu.cn).}
}

%
%

\markboth{Journal of \LaTeX\ Class Files,~Vol.~14, No.~8, December~2022}%
{Shell \MakeLowercase{\textit{et al.}}: Bare Demo of IEEEtran.cls for Computer Society Journals}
%



\IEEEtitleabstractindextext{%
\begin{abstract}
Representing visual data using compact binary codes is attracting increasing attention as binary codes are used as direct indices into hash table(s) for fast non-exhaustive search. Recent methods show that ranking binary codes using weighted Hamming distance (WHD) rather than Hamming distance (HD) by generating query-adaptive weights for each bit can better retrieve query-related items. However, search using WHD is slower than that using HD. One main challenge is that the complexity of extending a monotone increasing sequence using WHD to probe buckets in hash table(s) for existing methods is at least proportional to the square of the sequence length, while that using HD is proportional to the sequence length. To overcome this challenge, we propose a novel fast non-exhaustive search method using WHD. The key idea is to design a constant sequence extension algorithm to perform each sequence extension in constant computational complexity and the total complexity is proportional to the sequence length, which is justified by theoretical analysis. Experimental results show that our method is faster than other WHD-based search methods. Also, compared with the HD-based non-exhaustive search method, our method has comparable efficiency but retrieves more query-related items for the dataset of up to one billion items.
\end{abstract}

\begin{IEEEkeywords}
weighted Hamming distance, non-exhaustive search, hash table, binary codes, constant sequence extension
\end{IEEEkeywords}}

\maketitle

\IEEEdisplaynontitleabstractindextext

%
\IEEEpeerreviewmaketitle

\IEEEraisesectionheading{\section{Introduction}\label{sec:introduction}}

%
%
%
%

\IEEEPARstart{L}{arge-scale} vector search, which aims to retrieve query-related items from a large database of vectors efficiently and accurately is attracting increasing attention in visual retrieval~\cite{RN219,RN215,hu2022unsupervised,9244108,RN226,luo2020survey}. Since visual data are usually represented by high-dimensional vectors, conventional similarity search methods such as tree-based techniques for low-dimensional vectors are not suitable for visual data~\cite{RN219}. Therefore, it is desirable to develop similarity search methods for high-dimensional data.

Recently, hashing-based approximate nearest neighbor (ANN) search methods~\cite{RN219,RN214,RN211,lin2020hadamard,weng2020online,RN206,erin2015deep,RN201,RN217,RN212} have been developed to encode high-dimensional data into binary codes and perform fast non-exhaustive search on binary codes. Specifically, by learning similarity-preserving mappings~\cite{weiss2008spectral,wang2017robust,RN220}, hashing methods aim to generate compact binary codes to preserve Euclidean distance or semantic similarity of original data, and use these codes as direct indices (addresses) into hash table(s). Hence, given a query, hashing methods quickly retrieve related original data that are indexed by binary codes through probing their corresponding buckets of hash table(s).

\begin{figure*}[t]
	\centering
	\includegraphics[width=0.9\textwidth]{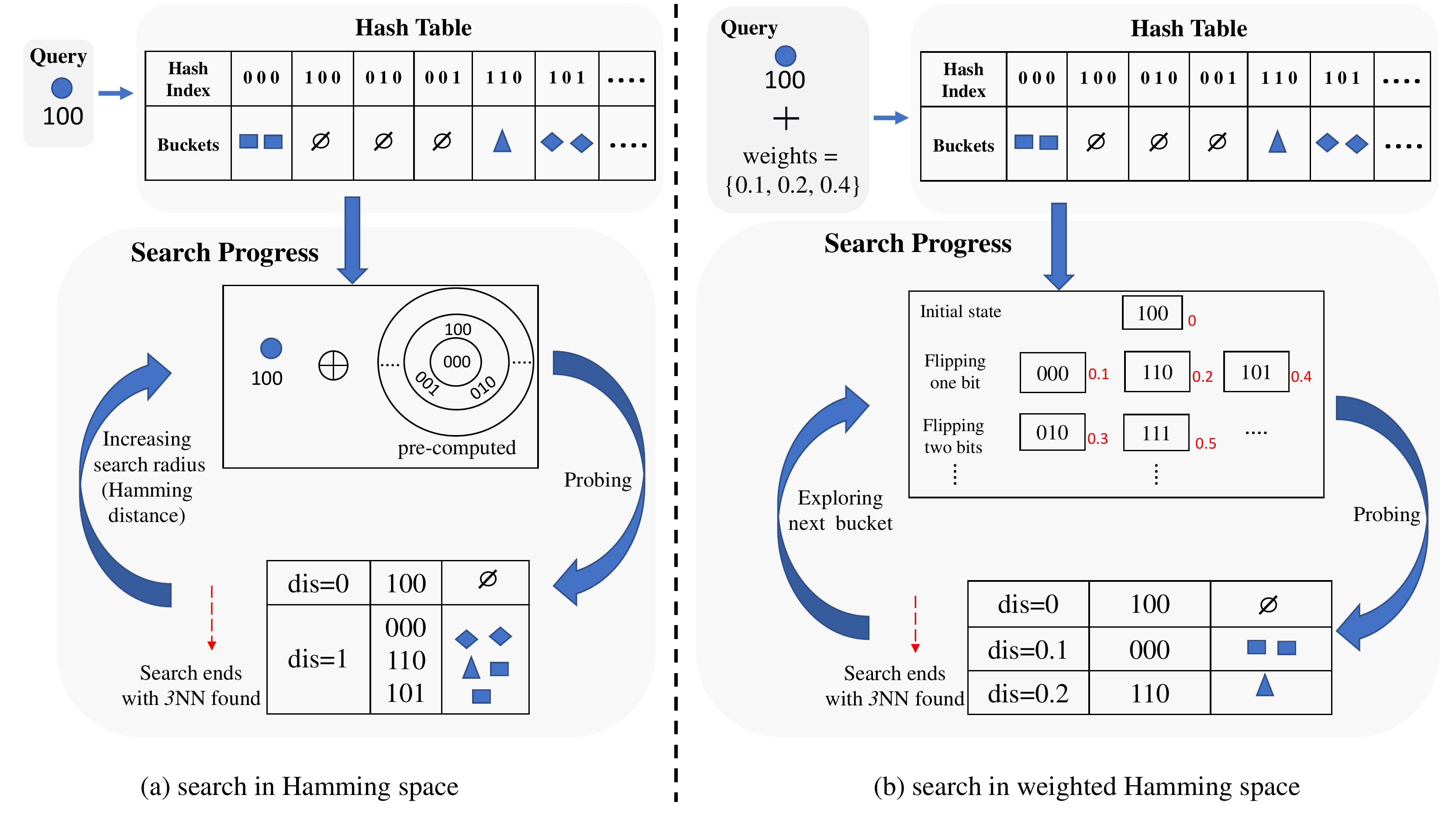}
	\caption{Search using Hamming distance (a) and weighted Hamming distance (b) using a hash table.}
	\label{fig:label1}

\end{figure*}

In search applications, binary codes are usually compared using Hamming distance (HD). But some studies~\cite{RN229,RN231} show that an ambiguity problem exists in HD calculation and affects the performance of binary codes. For example, in Fig.~\ref{fig:label1}(a), different binary codes have the same HD to a binary query, and it is difficult to discriminate which binary code is more related to the query in the original space than other binary codes. To address this issue, recent methods~\cite{RN228,RN232,RN234} consider generating bit-level weights and ranking binary codes using weighted Hamming distance (WHD) rather than HD. For example, the asymmetric distance (Asym) method~\cite{RN228} takes the data distribution of the database and the raw query point as the input and generates weights for each bit. By generating query-adaptive weights for each bit, different binary codes that have the same HD to the binary query can be discriminated, leading to improvement of the similarity-preserving performance of hashing methods.

However, performing non-exhaustive search on binary codes using WHD is much slower than that using HD. One main reason is that extending a monotone increasing sequence of binary codes with respect to WHD to probe buckets is much slower than that with respect to HD. For HD, a monotone increasing sequence starting with all bits set to zero can be customized in advance. Hence, a corresponding monotone increasing sequence for a given query is easily extended by performing an XOR operation between the pre-customized sequence and the query in constant computational complexity, which is shown in Fig.~\ref{fig:label1}. In contrast, the monotone increasing sequence with respect to WHD cannot be computed in advance since the weights depend on queries as described above. Existing non-exhaustive search methods~\cite{RN241,gui2021fast} using WHD explore the monotone increasing sequence by each time finding the smallest index ($i.e.,$ binary code) with WHD to the query among the generated candidate indices. What is worse, the number of generated candidate indices increases as the sequence length increases, and thus the computational complexity of each sequence extension is larger than that of the previous one. Hence, the complexity of the sequence extension for existing non-exhaustive search methods using WHD is at least proportional to the square of the sequence length.

In this paper, we propose a novel fast non-exhaustive search method using WHD. In the proposed method, we design a novel sequence extension algorithm, termed constant sequence extension (CSE), that can extend a monotone increasing sequence one by one in constant computational complexity. Specifically, for each sequence extension, we can find the smallest index with WHD to the query among a fixed number of candidate indices. Therefore, the total complexity of CSE is proportional to the sequence length, which is justified by theoretical analysis. By combining the CSE algorithm with a hash table, our search method can quickly probe buckets in the hash table one by one according to the elements in the sequence until $K$ nearest neighbors ($K$NNs) are found. Further, for long binary codes that require multiple hash tables to perform non-exhaustive search, our search method can directly integrate with multiple hash tables. The experimental results show that using either one or multiple hash tables, our method achieves dramatic speedups over a linear scan baseline and is faster than other non-exhaustive search methods using WHD. Especially, for the dataset of up to one billion items, compared with the non-exhaustive search using HD, the non-exhaustive search using WHD performed by the proposed method can retrieve more query-related items with comparable search speed.

The main contributions of this work are summarized as follows:
\begin{itemize}
	\item We propose a novel non-exhaustive search method on binary codes using WHD, which overcomes the high computational complexity problem of existing WHD-based search methods.
	\item We design a sequence extension algorithm, termed constant sequence extension, to perform each sequence extension in constant computational complexity, which is justified by theoretical analysis.
	\item In the experiments, compared with existing search methods using WHD, our method is faster and achieves same precision results for ANN search. Compared with the search method using HD, our method achieves higher precision results for ANN search with comparable efficiency.
\end{itemize}

The rest of this paper is organized as follows: We briefly review related works in Section 2. The proposed sequence extension algorithm is presented in Section 3 and the combination of the proposed sequence extension algorithm with hash table(s) is presented in Section 4. The experimental results are presented in Section 5. And finally, we conclude this paper in Section 6.

\section{Related Work}
\subsection{Weighted Hamming Distance}
Compared with HD, WHD can further discriminate different binary codes that have the same HD to the query and improve the performance of binary codes in search applications~\cite{RN232,RN234,RN230,RN227,RN235}. The weights for each bit are usually obtained by learning from data. For hashing methods, to improve the similarity-preserving performance, the asymmetric distance (Asym) method~\cite{RN228} takes the distribution of binary codes and the raw query point as the input and generates weights for each bit. It can be applied to different hashing methods~\cite{RN243,RN245}. The query-adaptive ranking (QRank) method~\cite{RN234} learns weights for each bit by exploiting the similarity relationship between the query point and its neighbors in the database. Some methods~\cite{RN231,shi2020discrete} utilize neural networks to generate binary codes and bit-level weights according to the query information and data information. Unlike the above methods that focus on learning weights for binary codes, our focus in this paper is on fast search for binary codes using WHD.

\subsection{Multi-Index Hashing}
By using binary codes as direct indices into a hash table, a non-exhaustive search is performed to quickly find $K$NNs and achieves a dramatic increase in speed over an exhaustive linear scan in terms of HD. However, the number of buckets in a hash table grows nearly-exponentially with the length of binary codes. When the length of binary codes is longer than 32 bits, the vast majority of buckets in a hash table will be empty as the number of buckets is much greater than the number of data points. Therefore, some methods resort to linear scan when the length of binary codes is longer than 32 bits. To address this issue, multi-index hashing (MIH)~\cite{RN233} builds multiple hash tables on binary code substrings. In detail, MIH separates binary codes into multiple disjoint substrings and builds multiple hash tables each of which takes the corresponding substrings as indices. By building multiple hash tables, non-exhaustive search can be performed on binary codes longer than 32 bits and achieve a dramatic speedup than linear scan.

In addition to HD, there are other similarity measures for binary codes in different applications~\cite{RN239,RN238}. To perform non-exhaustive search for binary codes with different similarity measures, various search methods based on hash table(s) are developed. The angular multi-index hashing (AMIH) method~\cite{RN239} is developed to search for binary codes with cosine similarity by using the connection between HD and cosine similarity. 

For WHD, the asymmetric distance (Asym) method~\cite{RN228} resorts to linear scan and accelerates the search by using lookup tables to store the weighted sums of binary codes that are computed in advance. However, it still performs exhaustive search. Inspired by MIH, non-exhaustive search on binary codes using WHD is performed by using multiple hash tables and exploring the monotone increasing sequence to probe buckets~\cite{gui2021fast,RN241}. The multiple tables in weighted hamming space (WHMT) method~\cite{gui2021fast} probes buckets to retrieve database items by extending a monotone increasing sequence. For each probing, WHMT generates some candidate indices and probes buckets according to candidate indices of which WHD is smaller than a threshold. The multi-index weighted querying (MIWQ) method~\cite{RN241} designs two operations to generate candidate indices for each probing and finds among all candidates the smallest one with WHD to the query to probe buckets. However, for both of these two methods, the number of generated candidate indices is more than one for each extension. With the increase of the length of a sequence, the number of generated candidate indices is accumulated and the computational complexity of finding the smallest one is increasing, which is time-consuming. In contrast, our method explores the monotone increasing sequence by extending each new element in constant computational complexity and thus is faster than the above search methods using WHD.

\begin{figure*}[t]
	\centering
	\includegraphics[width=0.95\textwidth]{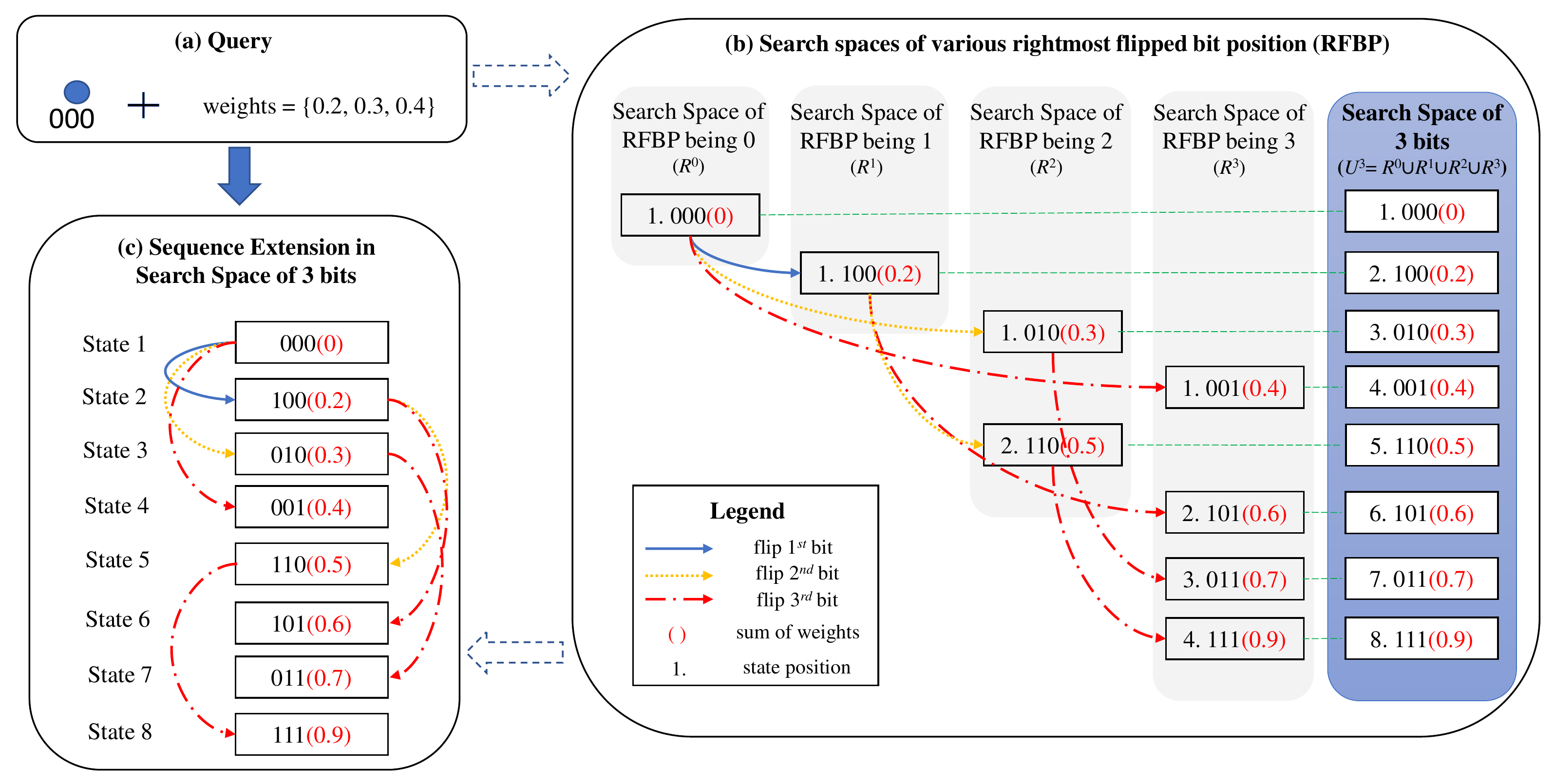}
	\caption{Analysis and framework of the proposed sequence extension algorithm. (a) gives the query and weights, (b) analyzes the relation between binary codes, and (c) is the framework of the proposed algorithm.}
	\label{fig:label2}

\end{figure*}

\section{Constant Sequence Extension}
In this section, we first define the problem of finding the smallest binary codes with WHD to the query in Sec. 3.1, which is shown in Fig.~\ref{fig:label2}(a). Then, we analyze how to find the smallest one among the remaining binary codes from previously found binary codes in Sec. 3.2, which is shown in Fig.~\ref{fig:label2}(b). According to the analysis, we design the constant sequence extension (CSE) algorithm in Sec. 3.3, which is shown in Fig.~\ref{fig:label2}(c). Finally, Sec. 3.4 elaborates the complexity analysis of the proposed algorithm and its implementation detail.

\subsection{Definition and Problem Statement}

Given a query $\bold{q}\in\{0,1\}^b$ where $b$ is the length of binary codes, HD between the binary code $\bold{g}$ and the query item $\bold{q}$ is defined as
\begin{equation}
	d_{H}(\mathbf{q},\mathbf{g})=\sum_{j=1}^{b}q_{j}\oplus \mathrm{g}_{j},
	\label{eqn1}
\end{equation}
where $\oplus$ is an XOR operation, $q_{j}$ is the $j^{th}$ bit of $\mathbf{q}$, and $\mathrm{g}_{j}$ is the $j^{th}$ bit of $\mathbf{g}$.

As the XOR result is 0 or 1 depending on whether the two bits are the same or not, HD ignores the distinction for each bit of binary codes. In contrast, WHD evaluates the distinction of each bit with a weight $w_{j}$ and is defined as 
\begin{equation}
	d_{w}(\mathbf{q},\mathbf{g})=\sum_{j=1}^{b}w_{j}(q_{j}\oplus \mathrm{g}_{j}),
	\label{eqn2}
\end{equation}
where $w_{j} \geq 0$. 

Given the query and weights, we aim to explore a monotone increasing sequence from the $b$-bit binary space consisting of $2^b$ states ($i.e.,$ binary codes). Specifically, the monotone increasing sequence is defined as $(\bold{h}_i)_{i=1}^L$ where $\bold{h}_i\in\{0,1\}^b$ is the $i^{th}$ smallest element with WHD to the query (the $i^{th}$ smallest element for short in the following) and $L$ is the length of the sequence. In the following, we propose a sequence extension algorithm to extend the target sequence efficiently.

\subsection{Sequence Extension Based on Previous States}
Given the query and weights, the query is fixed in the calculation of WHD between the query and each binary code. Therefore, we simplify the expression of Eq. (\ref{eqn2}) as 
\begin{equation}
	d_{w}(\mathbf{g})=d_{w}(\mathbf{q},\mathbf{g})
	\label{eqn3}
\end{equation}
where WHD between the query $\bold{q}$ and the binary code $\bold{g}$ is treated as a weighted sum of the binary code $\bold{g}$ with regard to the given query $\bold{q}$ and weights $w_j$.

Given the query and weights, the problem of finding the $i^{th}$ smallest element in the $b$-bit binary space can be decomposed into smaller sub-problems to formulate recurrence relations. The $b$-bit binary space is composed of $2^b$ binary codes whose length is $b$, and $\bold h_1 = \bold q$ is the smallest state in the $b$-bit space according to Eq. (\ref{eqn2}). The weighted sum of $\bold q$ is calculated as $d_{w}(\bold{q})=0$. When the $j^{th}$ bit of $\mathbf{h}_1$ is flipped ($i.e.$, from 0 to 1 or from 1 to 0), its weighted sum will increase by $w_j$. In the following, the weights $w_j$ are sorted from the smallest to the largest at the beginning. 

In Fig.~\ref{fig:label2}, we show a case of how to extend the sequence when the length of binary codes is 3 bits. From Fig.~\ref{fig:label2}(b), we can see that the sorted sequence is composed of the initial state and the states in the spaces of various rightmost flipped bit position (RFBP). Also, states in the search space of RFBP being $t$ are constructed by flipping the $t^{th}$ bit of states in the search space of RFBP being from 0 to $t-1$. Here, the search space of RFBP being 0 contains only one state, $\bold{h}_1$. 

Specifically, let $R^{t}$ denote the $b$-bit space with RFBP being $t$. That is, $R^{t}=\{\bold h \in \{0,1\}^b | {\bold h}[t] = 1\; {\rm and} \; {\bold h}[t+1..b] = 0 \}$, where ${\bold h}[t]=1$ and ${\bold h}[t]=0$ denote the $t^{th}$ bit of $\bold{h}_1$ is flipped and not flipped, respectively. Obviously, $R^{t} \subset \{0,1\}^b$ and the size of $R^{t}$ is $l(t) = 2^{(t-1)}$. Let $r^t(i)$ denote the weighted sum of the $i^{th}$ smallest state in $R^{t}$.
As $R^0$ only contains the initial state, $\bold{h}_1$, $r^0(1)=d_w(\bold{h}_1)$ and $l(0)=1$. From Fig.~\ref{fig:label2}(b), we can see that the monotone increasing sequence in $R^t$ is built by repeatedly appending into the sequence the smallest among the first remaining states from each monotone increasing sequence from $R^0$ to $R^{t-1}$ followed by flipping the $t^{th}$ bit. Let $U^{t}=R^0 \cup \cdots \cup R^t$ and $u^t(i)$ denote the weighted sum of the $i^{th}$ smallest state in $U^{t}$. The size of $U^{t}$ is $\hat l(t) = l(0) + \cdots l(t)= 2^t$. 

Assume the monotone increasing sequences from $R^0$ to $R^{t}$ are obtained. With iteration $i$ starting from 1 to $\hat l(t) = 2^{(t)}$, the weighted sum of the $i^{th}$ smallest state in $U^{t}$ is
\begin{equation}
	\begin{array}{c}
		u^t(i)=\min\limits_{j\in\{0,...,t\}\& p^j(i) \leq l(j)}  r^j(p^j(i)),  \\[4pt]
	\end{array}
	\label{eqn:dp1}
\end{equation}
where $p^j(i)$ point to the state position in $R^j$ when the $i^{th}$ smallest state in $U^{t}$ is extended and is initialized as $p^j(1)=1$ for $j \in \{0,...,t\}$. 

Then, the position pointer of the selected sequence moves to the remaining smallest state as 
\begin{equation}
	\begin{array}{c}
		j'=\argmin\limits_{j\in\{0,...,t\}\& p^j(i) \leq l(j)}  r^j(p^j(i)),  \\[4pt]
		p^{j}(i+1) = \left\{\begin{matrix} p^{j}(i) + 1 \;\;\;\;\;\;\;\;\; if \;\;\; j = j'
		\\ p^{j}(i) \;\;\;\;\;\;\;\;\;\;\;\;\;\; otherwise.
		\end{matrix}\right.		\\[4pt]
	\end{array}
	\label{eqn:pointer}
\end{equation}

With $\hat l(t) = 2^{t}$ iterations, we build the monotone increasing sequence in $U^{t}$. Then, with iteration $i$ starting from 1 to $l(t+1) = 2^{(t)}$, the weighted sum of the $i^{th}$ smallest state in $R^{t+1}$ is
\begin{equation}
	\begin{array}{c}
		r^{t+1}(i)= u^t(i) + w_{t+1}  \\[4pt]
	\end{array}
	\label{eqn:dp2}
\end{equation}

According to the above equations, the $i^{th}$ element in the target sequence we want to extend is the $i^{th}$ state in $U^{b}$, $i.e.,$ $d_w(\bold{h}_i)=u^{b}(i)$.

\subsection{Constant Sequence Extension}
According to Eq. (\ref{eqn:dp1}), we have $p^j(i) \leq i$ for each iteration since $i$ and $p^j(i)$ for $j \in \{0,...,b\}$ both start from 1. It means that if we want to find the $i^{th}$ smallest state in $U^b$, we just need to know the states preceding (including) the $i^{th}$ smallest state from $R^0$ to $R^b$. Therefore, we consider to obtain the $i^{th}$ smallest state in $U^b$ based on the preceding states as shown in Fig.~\ref{fig:label2}(c).

Combining Eq. (\ref{eqn:dp2}) with Eq. (\ref{eqn:dp1}), we have
\begin{equation}
	\begin{split}
		u^t(i)&=\min\limits_{j\in\{0,...,t\}\& p^j(i) \leq l(j)}  r^j(p^j(i))  \\[4pt]
	          &=\min\limits_{j\in\{0,...,t\}\& p^j(i) \leq l(j)}  u^{j-1}(p^j(i)) + w_j  \\[4pt]
	\end{split}
	\label{eqn:dp3}
\end{equation}

Considering the boundary case of $j=0$ in Eq. (\ref{eqn:dp3}), as we have $u^t(1)=u^0(1)=r^0(1)=0$ according to Eq. (\ref{eqn:dp1}), we reformulate Eq. (\ref{eqn:dp3}) below with $i$ starting from 2
\begin{equation}
		u^t(i)=\min\limits_{j\in\{1,...,t\}\& p^j(i) \leq l(j)}  u^{j-1}(p^j(i)) + w_j  \\[4pt]
	\label{eqn:dp4}
\end{equation}
where $p^j(1)=1$ for $j\in\{1,...,t\}$.

With weights sorted from smallest to largest, we can guarantee that $w^t(i) \geq w^{j-1}(p^j(i)) $ for $j\in\{1,...,t\}$ in each iteration $i$ in Eq. (\ref{eqn:dp4}) (see the proof of \textbf{Theorem 1}). Therefore, by constructing a function $f(j,k)$ (implementation detail will be described in the next subsection) to map the state position $k$ in $U^j$ ($j\leq b$) to the state position in $U^b$, we have
\begin{equation}
		u^b(i)=\min\limits_{j\in\{1,...,b\}\& p^j(i) \leq l(j)}  u^b(f(j-1,p^j(i))) + w_j  
	\label{eqn:dp5}
\end{equation}
where $u^b(1)=0$ and $i$ starts from 2.

According to Eq. (\ref{eqn:dp5}), the weighted sum of the $i^{th}$ state in $U^b$ is based on the preceding states in $U^b$. Also, the $i^{th}$ element in the target sequence we want to extend is the $i^{th}$ state in $U^b$, $i.e.,$ $d_w(\bold{h}_i)=u^{b}(i)$.

Correspondingly, we can obtain the monotone increasing sequence $(\bold{h}_i)_{i=1}^L$ where $\bold{h}_i\in\{0,1\}^b$ according to Eq. (\ref{eqn:dp5}). With $\bold{h}_1=\bold q$ and $i$ starting from 2, $\bold{h}_i$ is obtained by flipping the $j'^{th}$ bit of the $p^{j'}(i)^{th}$ element in the sequence in each iteration according to Eq. (\ref{eqn:dp5}). 

Therefore, the sequence extension for each iteration $i$ in our proposed CSE is formulated as
\begin{equation}
\begin{array}{c}	
	{u}^b(i)=\min\limits_{j\in\{1,...,b\}\& p^j(i) \leq l(j)}  u^b(f(j-1,p^j(i))) + w_j  \\[4pt]
	j'=\argmin\limits_{j\in\{1,...,b\}\& p^j(i) \leq l(j)}  u^b(f(j-1,p^j(i))) + w_j  \\[4pt] 
	\bold{h}_i = x(\bold{h}_{p^{j'}(i)}, j') \\[4pt]
	p^{j}(i+1) = \left\{\begin{matrix} p^{j}(i) + 1 \;\;\;\;\;\;\;\;\;\; if \;\; j = j'
	\\ p^{j}(i) \;\;\;\;\;\;\;\;\;\;\;\;\;\; otherwise.
	\end{matrix}\right.	
	\label{eqn:dparg5}
\end{array}
\end{equation}
where $x(\bold{h}, j)$ denotes flipping the $j^{th}$ bit of the binary code $\bold{h}$ and $b$ is the length of binary codes.

Before stating a main result in \textbf{Theorem 1}, we need \textbf{Lemma 1} first.

\textbf{Lemma 1} $u^j(i+1)-u^j(i) \leq  w_j$ for $j\in \{1,...,b\}$ and $i\in \{1,...,2^j-1\}$.
\begin{proof}
	It is proved by mathematical induction.
	
	For the base case, when $j=1$, there are only two values, $u^1(1)=0$ and $u^1(2)=w_1$. Obviously, $u^1(2)-u^1(1)=w_1$. Hence, the base case is correct. 
	
	For the induction step, suppose that $u^j(i+1)-u^j(i) \leq  w_j$ works from $j=1$ to $j=t$. According to Eq. (\ref{eqn:dp2}), we have $r^{t+1}(i+1)-r^{t+1}(i)\leq w_t$ for $i\in \{1,...,2^t-1\}$. Also, we have $r^{t+1}(i)-u^{t}(i)\leq w_{t+1}$ for $i\in \{1,...,2^t\}$. Therefore, as $w_{t+1} \geq w_t$, we have $u^{t+1}(i+1)-u^{t+1}(i) \leq  w_{t+1}$ for $i\in \{1,...,2^{t+1}-1\}$ according to Eq. (\ref{eqn:dp1}) and Eq. (\ref{eqn:dp2}). This concludes our proof.
	
\end{proof}

\textbf{Theorem 1} $u^t(i) \geq u^{j-1}(p^j(i)) $ for $j\in\{1,...,t\}$ in each iteration $i$ in Eq. (\ref{eqn:dp4}).
\begin{proof} 
It is proved by mathematical induction.
	
For the base case, $u^0(1)=u^j(1)=0$ for $j \in \{1,...,t\}$ in the initialization. As $i$ starts from 2, we have $u^t(2) \geq u^{j-1}(1)$ for $j \in \{1,...,t\}$ according to Eq. (\ref{eqn:dp4}).

For the induction step, suppose that $u^t(i) \geq u^{j-1}(p^j(i)) $ for $j \in \{1,...,t\}$ works for the iteration $i$. Also, assume $j$ is selected in Eq. (\ref{eqn:dp4}), and we have $u^t(i)=u^{j-1}(p^{j}(i)) + w_{j}$. According to \textbf{Lemma 1}, $u^{j-1}(p^{j}(i)+1)-u^{j-1}(p^{j}(i)) \leq w_{j-1}$. Hence, $u^t(i) \geq u^{j-1}(p^{j}(i)+1)$. Therefore, for the next iteration $i+1$, with $p^{j}(i+1)=p^{j}(i)+1$ according to Eq. (\ref{eqn:pointer}), as $u^t(i+1) \geq u^t(i)$, we have still $u^t(i+1) \geq u^{j-1}(p^j(i+1)) $ for $j \in \{1,...,t\}$. This concludes our proof.
\end{proof}

\begin{algorithm}[t]
	\caption{Constant Sequence Extension (CSE)} 
	\label{alg:dp} 
	\begin{algorithmic}[1] 
		\Require 
		query $\bold{q}$, weights $\{w_j\}_{j=1}^b$, code length $b$, $L$
		\Ensure 
		monotone increasing sequence $(\bold{h}_i)_{i=1}^L$
		\State Sort $\{w_j\}_{j=1}^b$ from smallest to largest
		\State Initialize $p(j)=1$, for $j=1,2,...,b$
		\State Initialize $l(j) = 2^j$, for $j=1,2,...,b$
		\State Initialize $\bold{h}_1 = \bold{q}$
		\For{$i$ = 2 to $L$}
		\State $t=-1$
		\State $v=\mathrm{Max}$
		\For{$j$=1 to $b$}
		\If{$p(j)\leq l(j)$ and $d_w(\bold{h}_{p(j)}) + w_j < v$}
		\State $t=j$
		\State $v=d_w(\bold{h}_{p(j)}) + w_j$
		\EndIf
		\EndFor
		\State $\bold{h}_{i}= x(\bold{h}_{p(t)}, t)$
		\State $p(t) += 1$
		\State MOD = $1<<t-1$
		\While{$p(t)\leq l(t)$ and $(\bold{h}_{p(t)} \oplus \bold{q})_{10} \%$ MOD != $(\bold{h}_{p(t)} \oplus \bold{q})_{10}$}
		\State $p(t) += 1$
		\State $l(t) += 1$
		\EndWhile
		\EndFor

	\end{algorithmic} 
\end{algorithm}

\begin{figure*}[t]
	\centering
	\includegraphics[width=0.9\textwidth]{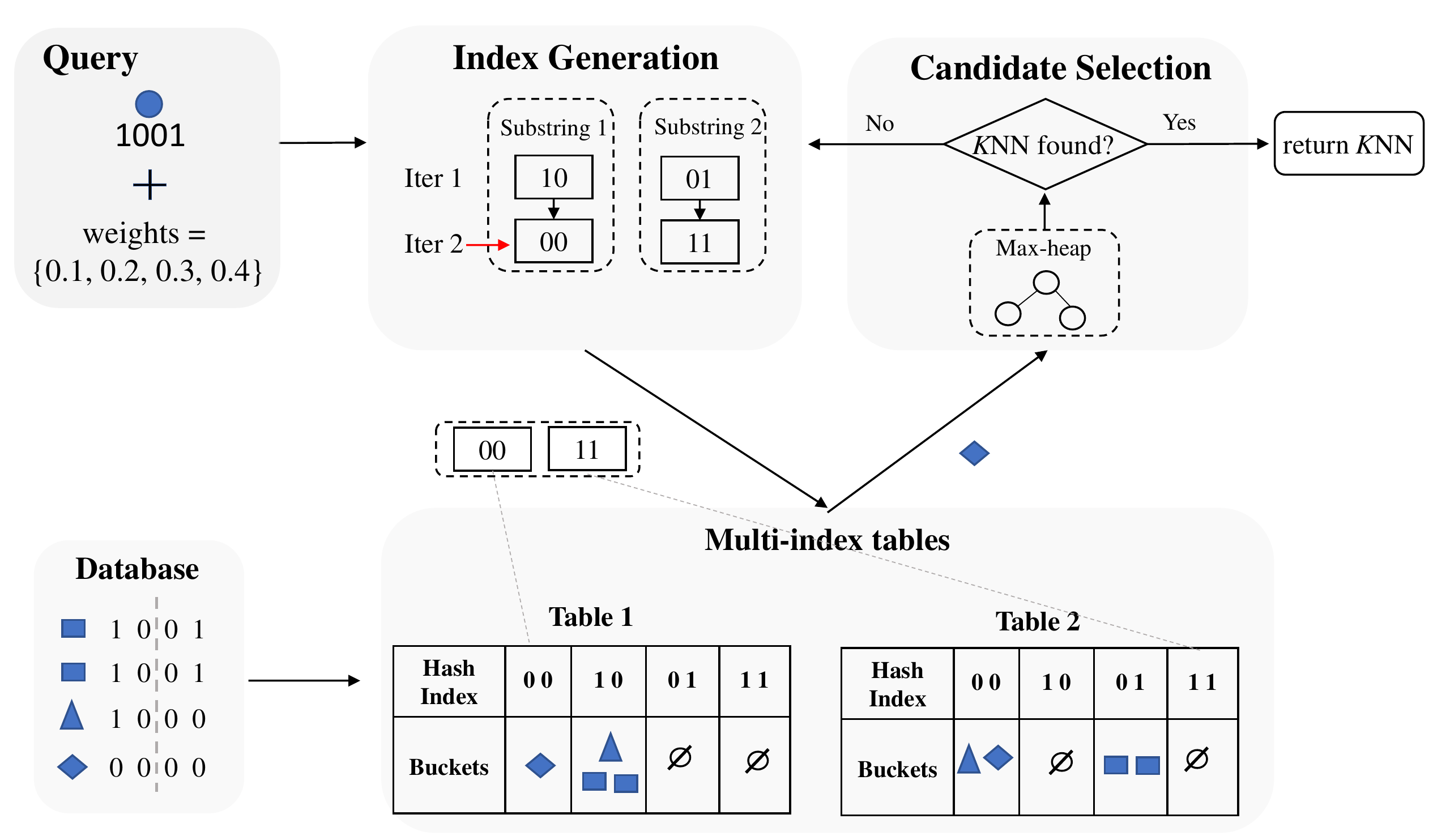}
	\caption{The proposed search framework with multi-index hash tables.}
	\label{fig:label3}

\end{figure*}

\subsection{Implementation Details and Complexity Analysis}
The pseudo-code of the proposed CSE algorithm is shown in Algorithm~\ref{alg:dp}. Particularly, steps 6-15 implement the recurrence of Eq. (\ref{eqn:dparg5}), and steps 16-20 implement the mapping function $f(j,k)$ to map the state position $k$ in $U^j$ ($j\leq b$) to the state position in $U^b$. $(\cdot)_{10}$ is to transform binary codes to decimal numbers. 

\textbf{Computational Complexity}: As shown in Algorithm~\ref{alg:dp}, the complexity of sorting weights is $O(b \log b)$ and the complexity of extending one element in the sequence is $O(b)$, which is independent of the sequence length. Hence, the total computational complexity of the proposed CSE is $O(b\log b+Lb)$ where $b$ is the length of binary codes and $L$ is the length of the sequence.

\textbf{Space Complexity}: As shown in Algorithm~\ref{alg:dp}, the proposed CSE mainly needs $O(Lb)$ to store the sequence of binary codes, $O(b)$ to store the weights, and $O(b)$ to store the state positions for each RFBP ($i.e.,$ $p(j)$). Hence, the total space complexity of the proposed CSE is $O(Lb)$.

\section{Fast Search with Weighted Hamming Distance}
In this section, we show how to perform $K$NN search using WHD by combining the proposed CSE with hash table(s). Following MIH~\cite{RN233}, we adopt one hash table for short binary codes and multiple hash tables for long binary codes, as described in Sec. 4.1 and Sec. 4.2, respectively.

\subsection{Search with One Hash Table}
To perform non-exhaustive $K$NN search using WHD based on a single hash table, we extend a monotone increasing sequence $(\bold{h}_i)$ to keep probing buckets until $K$NNs are found. Specifically, we keep generating new binary code according to Algorithm~\ref{alg:dp} and use that binary code to probe the corresponding bucket to retrieve the candidates. When we do not obtain $K$ candidates, we find the binary code that is the smallest one among all un-probed bucket indices ($i.e.,$ the smallest element among the remaining elements in the monotone increasing sequence) according to Algorithm~\ref{alg:dp}. Then, we probe the corresponding bucket to obtain the candidates. When we obtain $K$ candidates, these candidates are the $K$ nearest neighbors with WHD to the query. Algorithm~\ref{alg:one} shows our search method with one hash table. 
\begin{algorithm}[t]
	\caption{Search with One Hash Table} 
	\label{alg:one} 
	\begin{algorithmic}[1] 
		\Require 
		query $\bold{q}$, weights $\{w_j\}_{j=1}^b$, code length $b$, $K$
		\Ensure 
		set of $K$NNs $\Phi$ 
		\State Initialize $\Phi \leftarrow \emptyset$
		\State Adopt the initialization and $\bold{h}_1$ in Algorithm~\ref{alg:dp}
		\State Initialize $L=1$
		\While{$|\Phi| \le K$}
		\State Push candidates from bucket indexed by $\bold{h}_L$ into $\Phi$
		\State $L=L+1$
		\State Obtain $\bold{h}_L$ according to Algorithm~\ref{alg:dp}
		\EndWhile

	\end{algorithmic} 
\end{algorithm}

\subsection{Search with Multiple Hash Tables}

As demonstrated in MIH~\cite{RN233}, when the length $b$ of binary codes is larger than 32 bits, the number of buckets in a hash table will be very large ($i.e.,$ $2^b$), and the computational cost is extremely heavy. To mitigate this issue, MIH separates the binary codes into $m$ smaller disjoint substrings, and correspondingly creates a total of $m$ smaller hash tables by using each substring as index. These hash tables are called multi-index hash tables. However, the search method proposed in MIH is particularly designed for HD, and is not suitable for WHD~\cite{RN241,gui2021fast}. Different from~\cite{RN233}, our method can be directly applied to multiple hash tables to perform non-exhaustive $K$NN search using WHD.

As the binary codes are separated into $m$ smaller disjoint substrings, the length of each substring is $\left \lceil b/m \right \rceil$ or $\left \lfloor b/m \right \rfloor$. For convenience, we assume that $m$ is a factor of $b$, and that the substrings comprise continuous bits. The proposed search method based on multiple hash tables is shown in Fig.~\ref{fig:label3}, which is composed of two parts, index generation and candidate selection. The proposed search method performs multiple iterations to probe the buckets of each table until $K$NNs are found.

Specifically, in each iteration, the proposed method executes the index generation part and then the candidate selection part. In the index generation part, for each hash table, we extend a monotone increasing sequence of bucket indices by Algorithm 1 according to the weights of the corresponding substring. Hence, for each iteration, one bucket index of each hash table is generated and the corresponding bucket for each table is probed. That is, we extend the monotone increasing sequences $(\bold{h}^{(k)}_i)_{i=1}^{L}$, where $k\in\{1,...,m\}$ denotes the $k^{th}$ hash table and $L$ is the sequence length when $K$NNs are found according to Algorithm 1. For each iteration $i$, we probe the bucket of each hash table according to its corresponding index $\bold{h}_i^{(k)}$.

Next, with the candidates taken from the probed buckets of each table, the candidate selection part selects the $K$ nearest neighbors from the candidates. Following the literature~\cite{RN241}, we adopt a max-heap whose size is $K$ to implement candidate selection. The top node in the heap is the index with the maximum weighted sum. Specifically, after obtaining the candidates from the probed buckets, these candidates are pushed into the max-heap. When the first node of the max-heap is smaller than or equal to the weighted sum from the current probed bucket indices, the $K$ nearest neighbors are found. That is, the search ends when the first node of the max-heap $\bold{g}$ satisfies the condition below
\begin{equation}
d_w(\bold{g}) \leq \sum_{k=1}^{m}d_w^{(k)}({ \bold h}^{(k)}_i),
\label{eq:condition}
\end{equation}
where $d_w^{(k)}(\cdot)$ denotes the weighted sum of the $k^{th}$ substring.

\begin{algorithm}[t]
	\caption{Search with Multiple Hash Tables} 
	\label{alg:dp2} 
	\begin{algorithmic}[1] 
		\Require 
		query $\bold{q}$, weights $\{w_j\}_{j=1}^b$, code length $b$, $K$, $m$
		\Ensure 
		max-heap of $K$NNs $\Gamma$
		\State Initialize the max-heap $\Gamma$ of size $K$
		\State Obtain $\bold{h}^{(k)}_1$ according to $\bold{q}^{(k)}$ by Algorithm~\ref{alg:dp} for 
		\Statex $k \in \{1,...,m\}$
		\State Insert candidates from the bucket indexed by $\bold{h}^{(k)}_1$ of 
		\Statex the $k^{th}$ table into $\Gamma$ for $k \in \{1,...,m\}$
		\State $L$ = 1
		\While{not $\Gamma$.satisfied()}
		\State $L$ = $L$ + 1
		\State Obtain $\bold{h}^{(k)}_L$ according to $\bold{q}^{(k)}$ by Algorithm~\ref{alg:dp} for 
		\Statex \qquad $k \in \{1,...,m\}$
		\State Insert candidates from the bucket indexed by $\bold{h}^{(k)}_L$ of
		\Statex  \qquad the $k^{th}$ table into $\Gamma$ for $k \in \{1,...,m\}$
		\EndWhile

	\end{algorithmic} 
\end{algorithm}

Now we prove that our search method with multiple hash tables finds $K$NNs when the first node of the max-heap $\bold{g}$ satisfies the condition Eq. (\ref{eq:condition}) in \textbf{Theorem 2}.

\textbf{Theorem 2} When the first node of the max-heap $\bold{g}$ satisfies the condition Eq. (\ref{eq:condition}), the binary codes stored in the max-heap have the smallest sum of weights among all binary codes.
\begin{proof}
It is proved by contradiction. 

Assume there is a binary code $\bold{\hat g}$ that is not in the max-heap and is smaller than the first node of the max-heap $\bold{g}$. There exists $k \in \{1,...,m\}$ such that $d_w^{(k)}( \bold{\hat g}^{(k)}) \le d_w^{(k)}(\bold{g}^{(k)})$ where $\bold{\hat g}^{(k)}$ is the $k^{th}$ substring of $\bold{\hat g}$. Then, according to Eq. (\ref{eq:condition}), $d_w(\bold{\hat g}^{(k)}) \le d_w^{(k)}({\bold h}^{(k)}_i)$. Therefore, the binary code $\bold{\hat g}$ should have been taken out as a candidate from the bucket of the $k^{th}$ hash table according to Algorithm~\ref{alg:dp} and inserted into the max-heap. Obviously, the assumption is invalid. This concludes our proof.

\end{proof}

\begin{table*}[htbp]
	\setlength{\tabcolsep}{4pt}
	\centering
	\caption{Search time of different search methods on GIST1M. The search time is measured by millisecond. $m$ is the number of hash tables.}
	\begin{tabular}{|c|c|ccc|ccc|ccc|}
		\toprule
		\multirow{3}[6]{*}{Hashing-Weighting} & \multirow{3}[6]{*}{Method} & \multicolumn{9}{c|}{time cost (rounded speed-up factor of search method vs. linear scan) for $K$NN search} \\
		\cmidrule{3-11}          &       & \multicolumn{3}{c|}{32 bits ($m$=2)} & \multicolumn{3}{c|}{64 bits ($m$=4)} & \multicolumn{3}{c|}{128 bits ($m$=8)} \\
		\cmidrule{3-11}          &       & \multicolumn{1}{c|}{$K=1$} & \multicolumn{1}{c|}{$K=10$} & $K=100$ & \multicolumn{1}{c|}{$K=1$} & \multicolumn{1}{c|}{$K=10$} & $K=100$ & \multicolumn{1}{c|}{$K=1$} & \multicolumn{1}{c|}{$K=10$} & $K=100$ \\
		\midrule
		\multirow{5}[10]{*}{LSH-Asym} & \textbf{CSE} & \textbf{0.09(1254)} & \textbf{0.16(706)} & \textbf{0.4(282)} & \textbf{1.16(188)} & \textbf{2.37(92)} & \textbf{4.64(47)} & \textbf{11.68(38)} & \textbf{19.55(23)} & \textbf{30.76(15)} \\
		\cmidrule{2-11}          & WHMT  & 0.23(491) & 0.37(305) & 0.94(120) & 3.24(67) & 7.98(27) & 20.85(10) & 41.6(11) & 83.72(5) & 165.33(3) \\
		\cmidrule{2-11}          & MIWQ  & 0.11(1026) & 0.24(470) & 0.71(159) & 1.88(116) & 4.27(51) & 9.21(24) & 17.96(25) & 32.2(14) & 54.48(8) \\
		\cmidrule{2-11}          & Lookup & 24.15(5) & 24.15(5) & 24.15(5) & 41.45(5) & 41.45(5) & 41.45(5) & 81.27(5) & 81.27(5) & 81.27(5) \\
		\cmidrule{2-11}          & Linear Scan & 112.9(1) & 112.9(1) & 112.9(1) & 218.63(1) & 218.63(1) & 218.63(1) & 446.61(1) & 446.61(1) & 446.61(1) \\
		\midrule
		\multirow{5}[10]{*}{LSH-Qrank} & \textbf{CSE} & \textbf{0.07(1594)} & \textbf{0.15(744)} & \textbf{0.37(302)} & \textbf{1.17(185)} & \textbf{2.62(83)} & \textbf{4.78(45)} & \textbf{12.6(35)} & \textbf{21.7(20)} & \textbf{31.93(14)} \\
		\cmidrule{2-11}          & WHMT  & 8.19(14) & 11.72(10) & 13.86(8) & 27.24(8) & 32.87(7) & 38.48(6) & 72.16(6) & 88.12(5) & 103(4) \\
		\cmidrule{2-11}          & MIWQ  & 0.1(1116) & 0.23(485) & 0.67(167) & 2.01(108) & 4.89(44) & 9.21(24) & 18.64(24) & 33.81(13) & 51.32(9) \\
		\cmidrule{2-11}          & Lookup & 23.88(5) & 23.88(5) & 23.88(5) & 40.82(5) & 40.82(5) & 40.82(5) & 80.15(6) & 80.15(6) & 80.15(6) \\
		\cmidrule{2-11}          & Linear Scan & 111.56(1) & 111.56(1) & 111.56(1) & 216.71(1) & 216.71(1) & 216.71(1) & 442.04(1) & 442.04(1) & 442.04(1) \\
		\midrule
		\multirow{5}[10]{*}{ITQ-Asym} & \textbf{CSE} & \textbf{0.16(706)} & \textbf{0.23(491)} & \textbf{0.46(246)} & \textbf{1.05(208)} & \textbf{1.99(110)} & \textbf{3.76(58)} & \textbf{9.95(45)} & \textbf{15.9(28)} & \textbf{22.56(20)} \\
		\cmidrule{2-11}          & WHMT  & 0.35(323) & 0.51(222) & 0.92(123) & 2.5(88) & 5.48(40) & 12.72(17) & 30.3(15) & 60.38(7) & 122.17(4) \\
		\cmidrule{2-11}          & MIWQ  & 0.17(665) & 0.28(404) & 0.66(171) & 1.5(146) & 3.17(69) & 6.64(33) & 14.24(31) & 25.09(18) & 42.19(11) \\
		\cmidrule{2-11}          & Lookup & 24.12(5) & 24.12(5) & 24.12(5) & 41.4(5) & 41.4(5) & 41.4(5) & 81.24(6) & 81.24(6) & 81.24(6) \\
		\cmidrule{2-11}          & Linear Scan & 113.02(1) & 113.02(1) & 113.02(1) & 218.78(1) & 218.78(1) & 218.78(1) & 446.91(1) & 446.91(1) & 446.91(1) \\
		\midrule
		\multirow{5}[10]{*}{ITQ-Qrank} & \textbf{CSE} & \textbf{0.1(1115)} & \textbf{0.16(697)} & \textbf{0.36(310)} & \textbf{0.96(225)} & \textbf{1.96(110)} & \textbf{3.77(57)} & \textbf{10.41(42)} & \textbf{17.44(25)} & \textbf{26.33(17)} \\
		\cmidrule{2-11}          & WHMT  & 4.98(22) & 8.91(13) & 12.06(9) & 23.34(9) & 28.78(8) & 33.7(6) & 65.76(7) & 79.92(6) & 92.78(5) \\
		\cmidrule{2-11}          & MIWQ  & 0.11(1013) & 0.21(531) & 0.54(206) & 1.42(152) & 3.17(68) & 6.56(33) & 14.76(30) & 26.12(17) & 40.91(11) \\
		\cmidrule{2-11}          & Lookup & 23.83(5) & 23.83(5) & 23.83(5) & 40.9(5) & 40.9(5) & 40.9(5) & 80.31(5) & 80.31(5) & 80.31(5) \\
		\cmidrule{2-11}          & Linear Scan & 111.47(1) & 111.47(1) & 111.47(1) & 215.94(1) & 215.94(1) & 215.94(1) & 441.36(1) & 441.36(1) & 441.36(1) \\
		\bottomrule
	\end{tabular}%
	\label{tab:gist1m}%
\end{table*}%

Algorithm~\ref{alg:dp2} shows our search method with multiple hash tables, where $\Gamma$.satisfied() means that the first node of the max-heap $\bold{g}$ satisfies Eq. (\ref{eq:condition}).

\section{Experiments}
\subsection{Datasets and Settings}
In the experiments, we adopt three widely-used datasets for nearest neighbor search, SIFT1M, GIST1M, and SIFT1B, which are all from the BIGANN dataset~\cite{RN240}. The SIFT1M dataset contains 1 million 128-dimensional SIFT descriptors as base vectors and 10,000 queries. The GIST1M dataset contains 1 million 960-dimensional GIST descriptors as base vectors and 1,000 queries. The SIFT1B dataset is a much larger dataset, which contains 1 billion 128-dimensional SIFT descriptors as base vectors and 10,000 queries. The base vectors are used for search. 

We use two notable hashing methods to encode vectors into binary codes, namely, locality sensitive hashing (LSH)~\cite{RN243}, a data-independent hashing  method and iterative quantization (ITQ)~\cite{RN245}, a data-dependent hashing method. Also, we use two weighted methods to create weights for each bit of binary codes, namely, Asym~\cite{RN228} and QRank~\cite{RN234}. Their source codes are public.

\subsection{Comparison with WHD-based Search Methods}
In this subsection, we evaluate the search time of our method, fast search by constant sequence extension (CSE), and linear scan using WHD to verify the efficiency of CSE. Also, we compare CSE with other accelerating search methods using WHD. They are MIWQ~\cite{RN241}, WHMT~\cite{gui2021fast}, and Lookup~\cite{RN228}. Both MIWQ and WHMT use multiple hash tables to perform non-exhaustive search using WHD. Following MIWQ and WHMT, the number $m$ of hash tables for CSE, MIWQ, and WHMT in the search process is set as $m=\left[b/\log_2n\right]$~\cite{RN233}, where $b$ is the length of binary codes and $n$ is the number of database items. Lookup accelerates the exhaustive search using WHD by adopting the lookup tables to store the pre-computed weights for every eight bits. These methods are all implemented in
C++ and all the experiments are run on a single core Intel Core-i7 CPU with 64GB of memory.

Tables~\ref{tab:gist1m} and~\ref{tab:sift1m} show the average search time for each query of various search methods on GIST1M and SIFT1M, respectively. In the tables, the speed-up factor of various search methods vs. linear scan is defined as 
\begin{equation}
\textrm{speed-up factor} = \frac{\textrm{time (linear scan)}}{\textrm{time (search method)}}
\end{equation}
According to the results on GIST1M in Table~\ref{tab:gist1m}, CSE, WHMT, MIWQ, and Lookup are all faster than linear scan. CSE is the fastest method among them. Since Lookup performs exhaustive search, it is constantly about four times faster than linear scan for various lengths of binary codes. In contrast, since CSE, WHMT, and MIWQ perform non-exhaustive search, the speed-up factors of them against linear scan are different for various lengths of binary codes and various numbers of retrieved data points. They are faster when the length of binary codes is shorter and the number of retrieved data points is fewer. Especially in the case of retrieving one nearest neighbor for 32 bits, WHMT is at least ten times faster than linear scan, while CSE and MIWQ are at least hundred times faster. Besides, the binary codes and weights created by different methods also influence the speed-up factors of CSE, WHMT, and MIWQ against linear scan. For different binary codes and weights, CSE is much faster than linear scan and the fastest method among all the compared search methods.

\begin{table*}[htbp]
	\setlength{\tabcolsep}{2.5pt}
	\centering
	\caption{Search time of different search methods on SIFT1M. The search time is measured by millisecond. $m$ is the number of hash tables.}
	\begin{tabular}{|c|c|ccc|ccc|ccc|}
		\toprule
		\multirow{3}[6]{*}{Hashing-Weighting} & \multirow{3}[6]{*}{Method} & \multicolumn{9}{c|}{time cost (rounded speed-up factor of search method vs. linear scan) for $K$NN search} \\
		\cmidrule{3-11}          &       & \multicolumn{3}{c|}{32 bits ($m$=2)} & \multicolumn{3}{c|}{64 bits ($m$=4)} & \multicolumn{3}{c|}{128 bits ($m$=8)} \\
		\cmidrule{3-11}          &       & \multicolumn{1}{c|}{$K=1$} & \multicolumn{1}{c|}{$K=10$} & $K=100$ & \multicolumn{1}{c|}{$K=1$} & \multicolumn{1}{c|}{$K=10$} & $K=100$ & \multicolumn{1}{c|}{$K=1$} & \multicolumn{1}{c|}{$K=10$} & $K=100$ \\
		\midrule
		\multirow{5}[10]{*}{LSH-Asym} & \textbf{CSE} & \textbf{0.08(1411)} & \textbf{0.12(941)} & \textbf{0.29(389)} & \textbf{0.65(337)} & \textbf{1.26(174)} & \textbf{2.5(88)} & \textbf{4.66(96)} & \textbf{8.25(54)} & \textbf{13.8(32)} \\
		\cmidrule{2-11}          & WHMT  & 0.26(434) & 0.39(289) & 0.6(188) & 1.7(129) & 2.6(84) & 4.85(45) & 9.07(49) & 14.96(30) & 25.96(17) \\
		\cmidrule{2-11}          & MIWQ  & \textbf{0.08(1411)} & 0.13(868) & 0.36(314) & 0.72(304) & 1.49(147) & 3.15(70) & 5.19(86) & 9.45(47) & 16.57(27) \\
		\cmidrule{2-11}          & Lookup & 24.19(5) & 24.19(5) & 24.19(5) & 41.43(5) & 41.43(5) & 41.43(5) & 81.21(5) & 81.21(5) & 81.21(5) \\
		\cmidrule{2-11}          & Linear Scan & 112.89(1) & 112.89(1) & 112.89(1) & 219.12(1) & 219.12(1) & 219.12(1) & 445.13(1) & 445.13(1) & 445.13(1) \\
		\midrule
		\multirow{5}[10]{*}{LSH-Qrank} & \textbf{CSE} & \textbf{0.06(1835)} & \textbf{0.1(1101)} & \textbf{0.25(440)} & \textbf{0.59(361)} & \textbf{1.19(179)} & \textbf{2.44(87)} & \textbf{5.73(76)} & \textbf{9.74(45)} & \textbf{15.34(28)} \\
		\cmidrule{2-11}          & WHMT  & 2.48(44) & 6.67(17) & 11.24(10) & 18.73(11) & 25.35(8) & 30.68(7) & 51.7(8) & 63.35(7) & 74.42(6) \\
		\cmidrule{2-11}          & MIWQ  & 0.07(1572) & 0.12(917) & 0.33(334) & 0.73(292) & 1.59(134) & 3.48(61) & 6.36(69) & 11.06(39) & 18.25(24) \\
		\cmidrule{2-11}          & Lookup & 23.79(5) & 23.79(5) & 23.79(5) & 40.82(5) & 40.82(5) & 40.82(5) & 79.97(5) & 79.97(5) & 79.97(5) \\
		\cmidrule{2-11}          & Linear Scan & 110.07(1) & 110.07(1) & 110.07(1) & 213.2(1) & 213.2(1) & 213.2(1) & 436.25(1) & 436.25(1) & 436.25(1) \\
		\midrule
		\multirow{5}[10]{*}{ITQ-Asym} & \textbf{CSE} & \textbf{0.14(797)} & \textbf{0.19(588)} & \textbf{0.38(294)} & \textbf{0.76(284)} & \textbf{1.39(155)} & \textbf{2.58(84)} & \textbf{6.56(67)} & \textbf{10.73(41)} & \textbf{16.28(27)} \\
		\cmidrule{2-11}          & WHMT  & 0.42(266) & 0.62(180) & 0.85(131) & 1.95(111) & 2.63(82) & 4.14(52) & 10.71(41) & 15.91(28) & 25.34(17) \\
		\cmidrule{2-11}          & MIWQ  & \textbf{0.14(797)} & \textbf{0.19(588)} & 0.39(286) & 0.79(273) & 1.48(146) & 2.83(76) & 6.91(64) & 11.51(38) & 18.18(24) \\
		\cmidrule{2-11}          & Lookup & 24.15(5) & 24.15(5) & 24.15(5) & 41.42(5) & 41.42(5) & 41.42(5) & 80.76(5) & 80.76(5) & 80.76(5) \\
		\cmidrule{2-11}          & Linear Scan & 111.63(1) & 111.63(1) & 111.63(1) & 215.91(1) & 215.91(1) & 215.91(1) & 441.91(1) & 441.91(1) & 441.91(1) \\
		\midrule
		\multirow{5}[10]{*}{ITQ-Qrank} & \textbf{CSE} & \textbf{0.09(1240)} & \textbf{0.14(797)} & \textbf{0.29(385)} & \textbf{0.69(313)} & \textbf{1.36(159)} & \textbf{2.59(83)} & \textbf{6.5(68)} & \textbf{10.46(42)} & \textbf{15.62(28)} \\
		\cmidrule{2-11}          & WHMT  & 0.69(162) & 2.27(49) & 5.5(20) & 10.89(20) & 18.23(12) & 25.06(9) & 45.53(10) & 58.03(8) & 69.1(6) \\
		\cmidrule{2-11}          & MIWQ  & \textbf{0.09(1240)} & \textbf{0.14(797)} & 0.31(360) & 0.75(288) & 1.53(141) & 3.01(72) & 6.95(64) & 11.35(39) & 17.37(25) \\
		\cmidrule{2-11}          & Lookup & 23.81(5) & 23.81(5) & 23.81(5) & 40.86(5) & 40.86(5) & 40.86(5) & 79.85(6) & 79.85(6) & 79.85(6) \\
		\cmidrule{2-11}          & Linear Scan & 111.63(1) & 111.63(1) & 111.63(1) & 215.91(1) & 215.91(1) & 215.91(1) & 441.91(1) & 441.91(1) & 441.91(1) \\
		\bottomrule
	\end{tabular}%
	\label{tab:sift1m}%
\end{table*}%

\begin{table*}[htbp]
	\setlength{\tabcolsep}{2.5pt}
	\centering
	\caption{Search time of different search methods on SIFT1B. The search time is measured by millisecond. $m$ is the number of hash tables.}
	\begin{tabular}{|c|ccc|ccc|ccc|}
		\toprule
		\multirow{3}[6]{*}{Method} & \multicolumn{9}{c|}{time cost (speed-up factor of search method vs. linear scan) for $K$NN search} \\
		\cmidrule{2-10}          & \multicolumn{3}{c|}{32 bits ($m$=1)} & \multicolumn{3}{c|}{64 bits ($m$=2)} & \multicolumn{3}{c|}{128 bits ($m$=4)} \\
		\cmidrule{2-10}          & \multicolumn{1}{c|}{$K=1$} & \multicolumn{1}{c|}{$K=10$} & $K=100$ & \multicolumn{1}{c|}{$K=1$} & \multicolumn{1}{c|}{$K=10$} & $K=100$ & \multicolumn{1}{c|}{$K=1$} & \multicolumn{1}{c|}{$K=10$} & $K=100$ \\
		\midrule
		\textbf{CSE} & \textbf{8.21(13657)} & \textbf{8.22(13641)} & \textbf{8.24(13608)} & \textbf{9.16(23710)} & \textbf{11.45(18968)} & \textbf{19.03(11413)} & \textbf{43.16(10317)} & \textbf{87.33(5099)} & \textbf{200.54(2220)} \\
		\cmidrule{1-1}    WHMT  & 37.7(2974) & 66.18(1694) & 87.75(1278) & 1051.95(206)     & 6952.62(31)   & -     & -     & -     & - \\
		\cmidrule{1-1}    MIWQ  & \textbf{8.21(13657)} & 8.27(13558) & 8.3(13509) & 9.87(22004) & 14.26(15230) & 31.78(6834) & 94.32(4721) & 202.58(2198) & 531.69(837) \\
		\cmidrule{1-1}    Lookup & 23826.51(5) & 23826.51(5) & 23826.51(5) & 41693.37(5) & 41693.37(5) & 41693.37(5) & 82818.89(5) & 82818.89(5) & 82818.89(5) \\
		\cmidrule{1-1}    Linear Scan & 112128.01(1) & 112128.01(1) & 112128.01(1) & 217181.29(1) & 217181.29(1) & 217181.29(1) & 445283.36(1) & 445283.36(1) & 445283.36(1) \\
		\bottomrule
	\end{tabular}%
	\label{tab:sift1b}%
\end{table*}%

\begin{figure*}[t]
	\centering
	\includegraphics[width=0.85\textwidth]{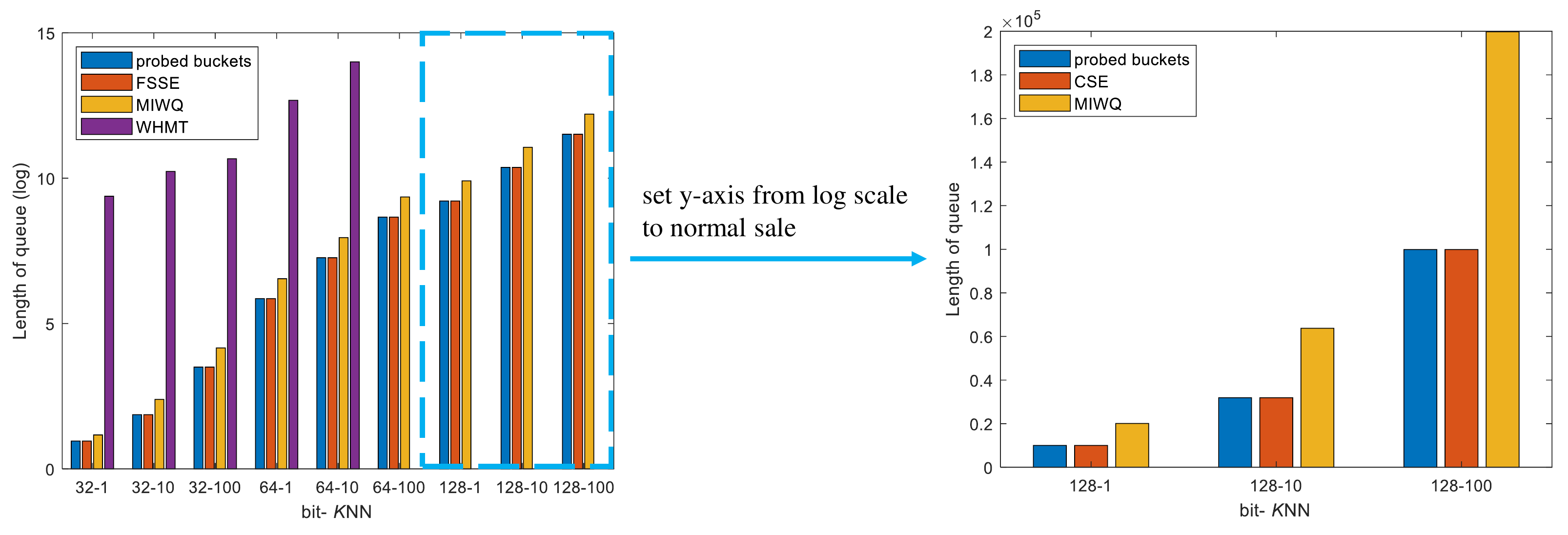}
	\caption{Length of candidate bucket index queue for CSE, MIWQ, and WHMT against number of probed buckets on SIFT1B.}
	\label{fig:MIWQ}
\end{figure*}

\begin{table*}[htbp]
	\setlength{\tabcolsep}{2.5pt}
	\centering
	\caption{Precision of different search methods on SIFT1B.}
	\begin{tabular}{|c|ccc|ccc|ccc|}
		\toprule
		\multirow{3}[6]{*}{Method} & \multicolumn{9}{c|}{precision@K(\%) for K-NN search } \\
		\cmidrule{2-10}          & \multicolumn{3}{c|}{32 bits} & \multicolumn{3}{c|}{64 bits} & \multicolumn{3}{c|}{128 bits} \\
		\cmidrule{2-10}         & \multicolumn{1}{c|}{$K=1$} & \multicolumn{1}{c|}{$K=10$} & $K=100$ & \multicolumn{1}{c|}{$K=1$} & \multicolumn{1}{c|}{$K=10$} & $K=100$ & \multicolumn{1}{c|}{$K=1$} & \multicolumn{1}{c|}{$K=10$} & $K=100$ \\
		\midrule
		\textbf{CSE} & 2.95 & 2.64 & 2.13 & 26.25 & 20.8 & 13.84 & 66.9 & 55.18 & 37.64 \\
		\cmidrule{1-1}    WHMT  & 2.95  & 2.64  & 2.13  & 26.25 & 20.8 & -     & -     & -     & - \\
		\cmidrule{1-1} MIWQ & 2.95 & 2.64 & 2.13 & 26.25 & 20.8 & 13.84 & 66.9 & 55.18 & 37.64 \\
		\cmidrule{1-1}    linear scan & 2.95  & 2.64  & 2.13  & 26.25 & 20.8  & 13.84 & 66.9  & 55.18 & 37.64 \\
		\bottomrule
	\end{tabular}%
	\label{tab:pre_M}%
\end{table*}%

\begin{table*}[htbp]
	\setlength{\tabcolsep}{2.5pt}
	\centering
	\caption{Precision comparison of CSE vs. MIH on SIFT1M and SIFT1B.}
	\begin{tabular}{|c|c|ccc|ccc|ccc|}
		\toprule
		\multirow{3}[6]{*}{} & \multirow{3}[6]{*}{Method} & \multicolumn{9}{c|}{precision@K(\%) for $K$NN search } \\
		\cmidrule{3-11}          &       & \multicolumn{3}{c|}{32 bits} & \multicolumn{3}{c|}{64 bits} & \multicolumn{3}{c|}{128 bits} \\
		\cmidrule{3-11}          &       & \multicolumn{1}{c|}{$K=1$} & \multicolumn{1}{c|}{$K=10$} & $K=100$ & \multicolumn{1}{c|}{$K=1$} & \multicolumn{1}{c|}{$K=10$} & $K=100$ & \multicolumn{1}{c|}{$K=1$} & \multicolumn{1}{c|}{$K=10$} & $K=100$ \\
		\midrule
		\multirow{2}[4]{*}{SIFT1M} & \textbf{CSE} & \textbf{32.14} & \textbf{28.42} & \textbf{21.59} & \textbf{69.77} & \textbf{59.86} & \textbf{44.79} & \textbf{93.58} & \textbf{86.43} & \textbf{69.91} \\
		\cmidrule{2-2}          & MIH   & 28.59 & 23.51 & 16.81 & 59.68 & 49.66 & 35.35 & 86.67 & 77.28 & 58.95 \\
		\midrule
		\multirow{2}[4]{*}{SIFT1B} & \textbf{CSE} & \textbf{2.95} & \textbf{2.64} & \textbf{2.13} & \textbf{26.25} & \textbf{20.8} & \textbf{13.84} & \textbf{66.9} & \textbf{55.18} & \textbf{37.64} \\
		\cmidrule{2-2}          & MIH   & 2.67  & 2.32  & 1.67  & 19.32 & 14.61 & 9.06  & 53.3  & 41.84 & 26.56 \\
		\bottomrule
	\end{tabular}%
	\label{tab:pre}%
\end{table*}%

\begin{table*}[t]
	\setlength{\tabcolsep}{2.5pt}
	\centering
	\caption{Search time comparison of CSE vs. MIH on SIFT1M and SIFT1B.}
	\begin{tabular}{|c|c|ccc|ccc|ccc|}
		\toprule
		\multirow{3}[6]{*}{} & \multirow{3}[6]{*}{Method} & \multicolumn{9}{c|}{time cost (speed-up factor of search method vs. linear scan) for $K$NN search} \\
		\cmidrule{3-11}          &       & \multicolumn{3}{c|}{32 bits} & \multicolumn{3}{c|}{64 bits} & \multicolumn{3}{c|}{128 bits} \\
		\cmidrule{3-11}          &       & \multicolumn{1}{c|}{$K=1$} & \multicolumn{1}{c|}{$K=10$} & $K=100$ & \multicolumn{1}{c|}{$K=1$} & \multicolumn{1}{c|}{$K=10$} & $K=100$ & \multicolumn{1}{c|}{$K=1$} & \multicolumn{1}{c|}{$K=10$} & $K=100$ \\
		\midrule
		\multirow{3}[6]{*}{SIFT1M} & \textbf{CSE} & 0.09(1254) & 0.16(706) & 0.4(282) & 1.16(188) & 2.37(92) & 4.64(47) & 11.68(38) & 19.55(23) & 30.76(15) \\
		\cmidrule{2-2}          & MIH   & 0.06(1882) & 0.08(1411) &0.17(664) & 0.32(685) & 0.59(371) & 1.13(194) & 1.71(260) & 3.03(147) & 5.21(85) \\
		\cmidrule{2-2}          & linear scan & 112.89(1) & 112.89(1) & 112.89(1) & 219.12(1) & 219.12(1) & 219.12(1) & 445.13(1) & 445.13(1) & 445.13(1) \\
		\midrule
		\multirow{3}[6]{*}{SIFT1B} & \textbf{CSE} & 8.21(13657) & 8.22(13641) & 8.24(13608) & 9.16(23710) & 11.45(18968) & 19.03(11413) & 43.16(10317) & 87.33(5099) & 200.54(2220) \\
		\cmidrule{2-2}          & MIH   & 8.09(13860) & 8.19(13691) & 8.3(13509) & 9.27(23428) & 11.51(18869) & 18.47(11759) & 183.05(2433) & 62.47(7128) & 135.76(3280) \\
		\cmidrule{2-2}          & linear scan & 112128.01(1) & 112128.01(1) & 112128.01(1) & 217181.29(1) & 217181.29(1) & 217181.29(1) & 445283.36(1) & 445283.36(1) & 445283.36(1) \\
		\bottomrule
	\end{tabular}%
	\label{tab:efficiency}%
\end{table*}%

The above analysis also holds for the results on SIFT1M in Table~\ref{tab:sift1m}. On SIFT1M, CSE is also the fastest method among all the compared search methods for various lengths of binary codes and various numbers of retrieved data points. It demonstrates that CSE is suitable for different data distributions. 

Further, we compare CSE with other methods on SIFT1B, which contains one billion vectors and is much larger than SIFT1M. Table~\ref{tab:sift1b} shows the results on SIFT1B, where the vectors are encoded into binary codes by LSH and the weights are created by Asym. Since WHMT requires a much longer candidate bucket index queue from 32 bits to 128 bits than MIWQ and CSE as shown in Fig.~\ref{fig:MIWQ}, WHMT is not available for 64 bits with $K=100$ and for 128 bits due to the memory limit. In Table~\ref{tab:sift1b}, CSE is still the fastest among all the search methods. Compared with the results in Table~\ref{tab:sift1m}, with the number of database vectors increasing, the speed-up factors of CSE, WHMT, and MIWQ become larger, while the speed-up factor of Lookup is still five. It demonstrates the advantage of performing non-exhaustive search by CSE, WHMT, and MIWQ. Compared with MIWQ which is the second fastest methods, the efficiency advantage of CSE becomes more obvious when the length of binary codes and the number of retrieved data points increase. It is because CSE can find $K$NNs with each sequence extension in constant computational complexity as analyzed in Sec. 3.3 while  
MIWQ with each sequence extension of which the complexity increases as the sequence length increases. Hence, CSE has a lower total complexity than MIWQ. 

Fig.~\ref{fig:MIWQ} shows the length of candidate bucket index queue for CSE, MIWQ, and WHMT against the number of probed buckets on SIFT1B. From the figure, we can see that when the length of binary codes and the number of retrieved data points increase, the number of probed buckets increases and thus the search time increases. The length of candidate bucket queue for CSE is the same as that of probed buckets as CSE generates one candidate index for each extension based on the previously found indices. In contrast, MIWQ and WHMT add bucket indices into the candidate queue and find the smallest one from the queue for each extension. The gap between the length of candidate queue for CSE and that for MIWQ is enlarged when the length of binary codes and the number of retrieved data points increase. Hence, when the length of binary codes and the number of retrieved data points increase, the efficiency advantage of CSE becomes more obvious compared with MIWQ. 

Table~\ref{tab:pre_M} shows the precision results for approximate nearest neighbor (ANN) search among database vectors on SIFT1B. Following the dataset setting~\cite{RN240}, the ground truth (GT) is defined as the 1,000 nearest neighbors among the base vectors according to the Euclidean distance for each query in the original space. $\textrm{precision@}K = \textrm{num (GT)}/K$ where num (GT) denotes the number of GT nearest neighbors of the query in the retrieved $K$ data points. According to the results in Table~\ref{tab:pre_M}, CSE, WHMT, MIWQ, and linear scan achieve the same precision results. It demonstrates that CSE, MIWQ, and WHMT can perform the exact non-exhaustive search for binary codes using WHD.

\subsection{Comparison with MIH}
We compare CSE with MIH~\cite{RN233}, which performs non-exhaustive search based on HD. In the following, we will compare CSE with MIH with respect to search precision and efficiency.

Table~\ref{tab:pre} shows the precision results of CSE vs. MIH for ANN search among database vectors on SIFT1M and SIFT1B. The database vectors in both datasets are encoded into binary codes by LSH and weights are created by Asym. According to the results on SIFT1M in Table~\ref{tab:pre}, as CSE performs search based on WHD and MIH based on HD, CSE can achieve higher precision results than MIH. With the length of binary codes increasing, both methods achieve higher precision results. By comparing the results on SIFT1B with those on SIFT1M, we can find that the difficulty of retrieving GT neighbors increases for both methods with the number of database vectors increasing. CSE is still better than MIH. Especially, with the length of binary codes increasing, the advantage of CSE against MIH is expanding on SIFT1B. It demonstrates the precision advantage for ANN search by searching with WHD over HD.

\begin{figure}[t]
	\centering
	\includegraphics[width=0.45\textwidth]{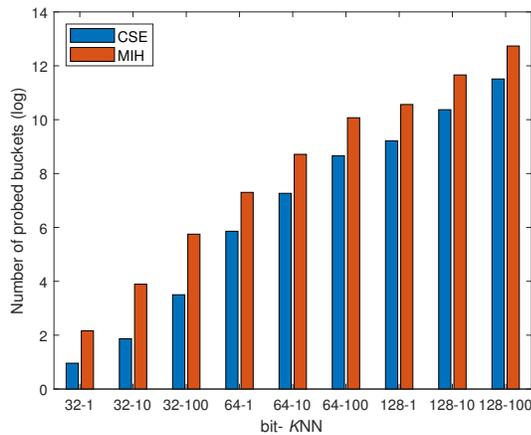}
	\caption{Number of probed buckets for CSE and MIH on SIFT1B.}
	\label{fig:MIH}
\end{figure}

Table~\ref{tab:efficiency} shows the average search time for each query of various search methods on SIFT1M and SIFT1B. According to the results on SIFT1M in Table~\ref{tab:efficiency}, MIH is obviously faster than CSE since the computation of HD is faster than that of WHD. However, CSE can achieve comparable efficiency with MIH on SIFT1B which contains much larger database vectors than SIFT1M. It is because with the number of database vectors increasing, the search time cost depends more on buckets probing than distance computation. Specifically, as shown in Fig.~\ref{fig:MIH}, the number of probed buckets of CSE is much smaller than MIH, so that the efficiency of probing buckets can compensate the inefficiency of computing WHD for CSE. The reason why CSE can probe fewer buckets than MIH may be that with the weights as the guide, CSE can probe the more similar and non-empty buckets than MIH according to the query. According to the results in Tables~\ref{tab:pre} and~\ref{tab:efficiency}, on SIFT1B, CSE has the obvious precision advantage with comparable efficiency compared with MIH for the dataset of up to one billion items.

\section{Conclusion}
In this paper, we have proposed a novel fast non-exhaustive search method using weighted Hamming distance (WHD). To solve the high computational complexity problem for each sequence extension in existing WHD-based search methods, we design a sequence extension algorithm, termed constant sequence extension, to perform each extension in a constant complexity. It is justified by theoretical analysis. The experiments show that our method is faster than other search methods using WHD. Also, compared with the non-exhaustive search using Hamming distance (HD), the non-exhaustive search using WHD performed by our method can achieve higher precision results for approximate nearest neighbor search with comparable efficiency for the dataset of up to one billion items. It shows that our method has the potential of having comparable efficiency compared with search using HD for large-scale data and will advance the development of weighted methods to learn weights for different bits in various search applications.


%





\ifCLASSOPTIONcaptionsoff
  \newpage
\fi



\bibliographystyle{IEEEtran}
\bibliography{egbib}

\end{document}